\documentclass[aps,prl,superscriptaddress,twocolumn,tightenlines]{revtex4}

\usepackage {graphicx}
\usepackage {amsmath}
\usepackage {amsthm}
\usepackage {amsfonts}
\usepackage {amssymb}
\usepackage {mathrsfs}

\usepackage{epstopdf}

\newcommand {\be}{\begin{eqnarray}}
\newcommand {\ee}{\end{eqnarray}}

\begin{document}

\title{Quantum Ferroelectric Magnets in Underdoped Manganites}

\author {E.R.\ S\'anchez Guajardo}
\email {edmundo@lorentz.leidenuniv.nl}
\affiliation {Instituut-Lorentz, Universiteit Leiden, P.O.\ Box 9506, 2300 RA Leiden, 
The Netherlands}

\date{\today}

\begin{abstract}
Using the degenerate double exchange Hamiltonian with on-site Coulomb interactions we show that there is an instability towards the displacement of the manganese ion in La$_{1-x}$Ca$_{x}$MnO$_{3}$.  The result is a dipole moment due to the charge disproportionation and the lattice distortions that gives rise to the predicted magnetic ferroelectric phase for dopings in between $x=4$ and $x=5$~\cite{jvdbnat}.  The instability is stabilized by the phonons of the lattice, resulting in a stable configuration of the degrees of freedom of the system.
\end{abstract}

\maketitle

Recently, manganites R$_{1-x}$A$_{x}$MnO$_{3}$ (R=rare-earth ion, A=divalent substituent), with their particular colossal magnetoresistance, have been the subject of intense study~\cite{jvdbnat, feiner, exp, little, orla, graaf, grenier,coinf}.  The internal degrees of freedom, charge, orbital state, and spin, vary with the composition $x$, resulting in different properties.  Interestingly, these compounds very rarely are both magnetic and ferroelectric, which would be very desirable for technological applications.

It is well known that ferroelectric perovskite compounds originate in the displacement of the mostly nonmagnetic transition metal ion from the center of the octahedron~\cite{cohen}.  The result is a stronger covalent bond between the transition metal ion and one of the surrounding oxygens.  The existence of a ferroelectric state in underdoped manganites due to the charge order that results in a lack of inversion symmetry has also been shown~\cite{jvdbnat}.  We show that the lattice distortions due to the displacement of the magnetic manganese ion, in addition to the previously discussed  interaction between charge, orbital, and spin ordering in underdoped manganites~\cite{jvdbprl,jvdbnat}, further lower the energy of the ferroelectric state, and give rise to an electric dipole moment.  The result is the desired ferroelectric magnet compound.

The main complication arises from the anisotropy of the orbital states of the manganese ion.  The fivefold degeneracy of its $d$ orbital states is partially lifted by the crystal field into twofold degenerate orbital states, d$_{x^{2}-y^{2}}$ and d$_{3z^{2}-r^{2}}$, called $e_{g}$, and threefold degenerate orbital states, d$_{xy}$, d$_{yz}$, and d$_{zx}$, called $t_{2g}$.  The degeneracy of the $e_{g}$ states is further lifted by the lattice distortions, or by the on-site Coulomb interactions. The former lift the degeneracy completely by lowering the symmetry of the cubic crystal which lowers the energy of the  d$_{3z^{2}-r^{2}}$ state while raising that of  d$_{x^{2}-y^{2}}$.  The latter make the system quasidegenerate because the degeneracy is only lifted after either of the two orbital states is first occupied.  After choosing an orbital basis we may characterize the orbital states of every ion by an angle~\cite{feiner}.

The electronic configuration of each ion in the array may also vary.  The Mn$^{3+}$ ion in the parent compound of interest, LaMnO$_{3}$, has an open shell configuration in which three electrons are in the $t_{2g}$ orbitals and one in either of the two $e_{g}$ orbitals.   The $t_{2g}$ electrons are localized due to a stronger stabilization of the crystal field with respect to the $e_{g}$ orbitals and a lower hybridization with the oxygens 2p states~\cite{tokura}.  When doping with a divalent ion that introduces holes, for instance La$_{1-x}$Ca$_{x}$MnO$_{3}$, some sites change to Mn$^{4+}$ with a closed shell configuration in which the $e_{g}$ orbitals are empty.

\begin{figure} [tb]
\includegraphics[width=0.39\textwidth]{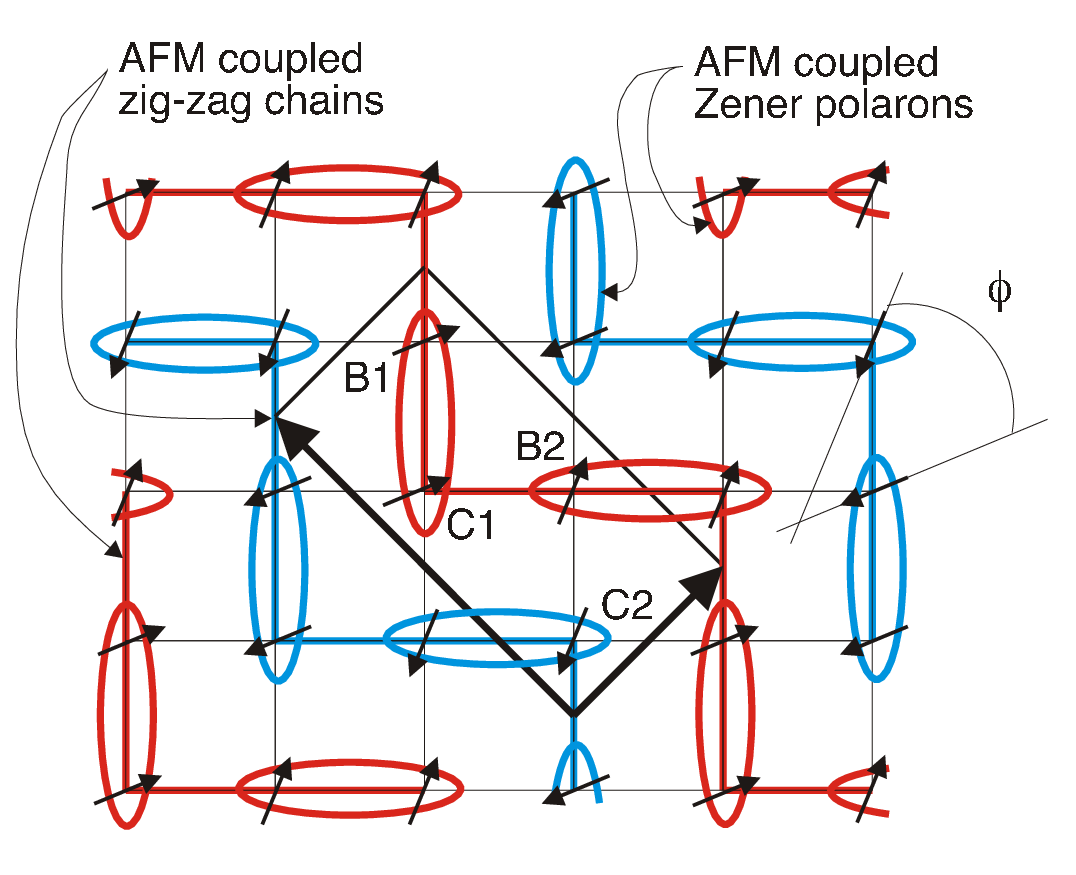}
\caption{\label{unit}The lattice with the corresponding unit cell for the intermediate phase $\phi=\pi/4$.  The CE phase corresponds to $\phi=0$, and the Zener polaron to $\phi=\pi/2$.  The Zener polaron phase is made up of dimers, with blue and red dimers AFM coupled within a diagonal.  The CE phase is composed of AFM coupled zig-zag chains.  The intermediate phase shows the superposition of the Zener polaron and the CE phase.  }
\end{figure}

There are three interactions that reproduce the properties of the solids in discussion:  the double-exchange, the superexchange, and the on-site Coulomb interaction.  The double-exchange, associated with the hopping amplitude $t$, leads to the formation of bands and to the conduction of electrons.  The superexchange, with the energy scale $J$, is a magnetic interaction between localized spins of neighboring sites.  The on-site Coulomb interaction is between electrons in different $e_{g}$ orbital states in the same ion.  Its energy scale is  $U$, and in the solids of interest we find $U>>t$, which tends to localize the electrons.

We consider two limiting cases of La$_{1-x}$Ca$_{x}$MnO$_{3}$:  the Zener polaron phase at $x=0.4$ doping, and the CE phase at $x=0.5$.  Both phases are known to be insulating.  As shown in Fig.~\ref{unit}, the CE phase is made of zig-zag chains with ferromagnetic spin order within the chain, and that are antiferromagnetically coupled with the neighboring chains.  This phase has the checkerboard charge order in which sites alternate between Mn$^{3+}$ and Mn$^{4+}$.  The unit cell has four sites, with one Mn ion each. The orbital order is d$_{3z^{2}-r^{2}}$ and d$_{x^{2}-y^{2}}$ in the corner sites (C1 and C2), d$_{3x^{2}-r^{2}}$ in the horizontal bridge (B2), and d$_{3y^{2}-r^{2}}$ in the vertical bridge (B1).  The Zener polaron phase is made up of dimers with equivalent sites (B1=C1 and B2=C2).  The spins within a dimer are parallel, while spins in between neighboring dimers with the same crystallographic direction couple antiferromagnetically.  Perpendicular dimers have perpendicular spin orientations with respect to each other.  The charge in a dimer is delocalized in between the equivalent sites, which results in a bond-centered charge order.  

The Hamiltonian of the system is given by
\begin{eqnarray}
H&=&t\sum_{<ij>,\sigma\sigma^{\prime}} \left<\chi_{i}|\chi_{j}\right>c_{i,\sigma}^{\dagger}c_{j,\sigma^{\prime}}\Gamma_{ij}^{\sigma\sigma^{\prime}} \nonumber\\
& &{  }+J_{AF}\sum_{<ij>}\vec{S}_{i}\cdot\vec{S}_{j} + U\sum_{i,\sigma\neq\sigma^{\prime}} n_{i}^{\sigma}n_{i}^{\sigma^{\prime}},
\label{Hamiltonian}
\end{eqnarray}
where $i,j$ are site labels, and $\sigma,\sigma^{\prime}$ denote $e_{g}$ orbital states. The sum $<ij>$ is over nearest neighbor sites. The on-site Coulomb interaction contains the density per $e_{g}$ orbital per site, $n_{i}^{\sigma}$, and $J_{AF}$ is the antiferromagnetic superexchange interaction between (classical) localized $t_{2g}$ spins, $\vec{S}_{i}$.  The kinetic term with the hopping integral $t$ contains the fermionic creation and annihilation operators, $c_{i,\sigma}^{\dagger}$ and $c_{j,\sigma^{\prime}}$, and the symmetry matrix elements $\Gamma^{\sigma\sigma^{\prime}}_{ij}$ that take into account the overlap of the orbitals involved and the crystallographic direction of the hopping.  We take into account the overlap of the spin wave functions~\cite{nagaosa}, $\left<\chi_{i}|\chi_{j}\right>$, which depends on the relative direction between the two spins, $|\left<\chi_{i}|\chi_{j}\right>|=\cos(\phi_{ij}/2)$.  For parallel spins the hopping is largest, and antiparallel spins cannot hop.

We consider the zero temperature limit with strong ferromagnetic Hund's coupling between localized $t_{2g}$ and itinerant $e_{g}$ electrons, $J_{H}/t\rightarrow\infty$.  The antiferromagnetic interaction becomes  $J_{AF}\sum_{<ij>}|S|^{2}\cos{\phi_{ij}}$, where the angle $\phi_{ij}$ denotes the relative orientation between spins.  The localized spins have $|S|^{2}=9/4$.  With respect to the symmetry matrix, we choose our orbital basis states to be $\alpha=3z^{2}-r^{2}$, and $\beta=x^{2}-y^{2}$.  In that case we have
\begin{eqnarray}
\Gamma_{<ij>//y}=\frac{1}{4}\left[
\begin{array} {cc}
1 &\sqrt{3}\\
\sqrt{3} &3
\end{array}
\right] ,
\label{gamma}
\end{eqnarray}
that takes into account the overlaps of the $e_{g}$ orbitals involved when hopping along the $y$-axis~\cite{dagotto}.  The hopping along the $x$-axis has a phase with respect to the $y$-axis which results in $\Gamma^{\alpha\beta}_{<ij>//x}=\Gamma^{\beta\alpha}_{<ij>//x}=-\sqrt{3}$.  After choosing a basis we may refer the orbital states of every ion, $\alpha_{i}$ and $\beta_{i}$, to the reference basis with the transformation
\begin{eqnarray}
\left[
\begin{array} c
|\alpha_{i}(\xi)>\\
|\beta_{i} (\xi)>
\end{array}
\right]&=& \left[
\begin{array} c
\cos(\frac{\xi}{2})|\alpha > + \sin(\frac{\xi}{2})|\beta >\\
-\sin(\frac{\xi}{2})|\alpha> + \cos(\frac{\xi}{2})|\beta>
\end{array}
\right].
\label{rotation}
\end{eqnarray}
Finally, we take the on-site Coulomb interaction to be $U/t=4$ and use the mean-field approximation of the densities $<n_{i}^{\sigma}>_{mf}=<c_{i,\sigma}^{\dagger}c_{i,\sigma}>$.

After decoupling the on-site Coulomb interaction, $n_{i}^{\alpha}n_{i}^{\beta}\approx <n_{i}^{\alpha}>n_{i}^{\beta}+n_{i}^{\alpha}<n_{i}^{\beta}>-<n_{i}^{\alpha}>< n_{i}^{\beta} >$, we first calculate the charge order (CO), orbital order (OO), and magnetic order (SO) for different fixed values of the doping varying the spin canting angle from $\phi_{ij}=0$ (CE phase) to $\phi_{ij}=\pi/2$ (Zener polarons).  Since we consider the relative orientation between spins, $\phi_{ij}$, to change equally, in what follows we use a general spin canting angle, $\phi$.  We also consider that the relative angle does not change for some neighboring spins, as seen in Fig.~\ref{unit}.  The anisotropy of the system is taken into account by rotating the orbital states per site to our working bases, $\alpha$ and $\beta$, with the transformation
\begin{eqnarray}
n_{i}^{\alpha}&=&n_{i}^{\alpha_{i}}\sin^{2}(\xi_{i}/2)+n_{i}^{\beta_{i}}\cos^{2}(\xi_{i}/2)\nonumber\\
n_{i}^{\beta}&=&n_{i}^{\alpha_{i}}\cos^{2}(\xi_{i}/2)+n_{i}^{\beta_{i}}\sin^{2}(\xi_{i}/2)
\label{rotation}
\end{eqnarray}
before solving numerically.   The self consistency equations,
\begin{eqnarray}
\xi_{i}&=&\tan^{-1}(2<c_{i,\alpha}^{\dagger}c_{i,\beta}>/(\tilde{n}_{i}^{\alpha}-\tilde{n}_{i}^{\beta}))\\
n_{i}^{\alpha_{i}}&=&\tilde{n}_{i}^{\alpha}\cos^{2}(\xi_{i}/2)+<c_{i,\alpha}^{\dagger}c_{i,\beta}>\sin(\xi_{i})\nonumber\\
&&{  }+\tilde{n}_{i}^{\beta}\sin^{2}(\xi_{i}/2)\nonumber\\
n_{i}^{\beta_{i}}&=&\tilde{n}_{i}^{\alpha}\sin^{2}(\xi_{i}/2)-<c_{i,\alpha}^{\dagger}c_{i,\beta}>\sin(\xi_{i})\nonumber\\
& &{  }+\tilde{n}_{i}^{\beta}\cos^{2}(\xi_{i}/2),\nonumber
\label{selfconsistent}
\end{eqnarray}
enforce that the rotation of the calculated densities, $\tilde{n}_{i}^{\sigma}$, from the working basis back to the on-site basis results in a diagonal on-site Coulomb interaction.  

\begin{figure}[tb]
\includegraphics[width=0.48\textwidth]{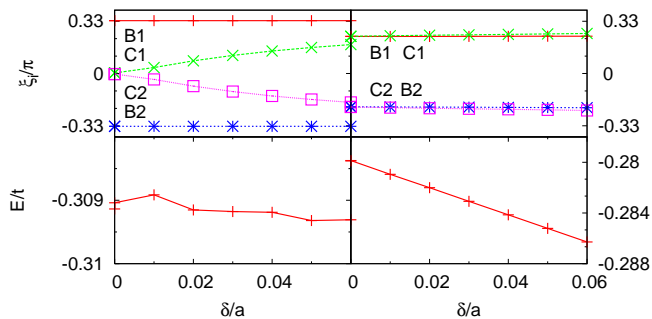}
\caption{\label{DistLim}Orbital order for fixed doping (top row), and energy without elastic energy contribution (bottom).  The left column corresponds to the CE phase, and the right to the Zener polaron phase.}
\end{figure}

Now comes the totally new result: we also take into account the displacement of the manganese ions by considering small changes in the hopping amplitude, $t\longrightarrow t+ c\delta$, where $\delta$ is the distortion of the lattice and the constant $c=3.5$ is obtained from Ref.~\cite{dagottol}.  We first fix the spin canting while varying $\delta$.  Regarding the orbital states of the CE phase, shown in Fig.~\ref{DistLim}, the effect of the distortions is to rotate the orbital states at the corner sites towards the bridge site orbital states making them more equivalent.  For the equivalent orbital states of the Zener polaron phase there is no change, which shows that the orbitals are independent of the distortion.  Considering the energy up to linear order in $\delta$ and without the elastic energy quadratic in $\delta$ we find an almost constant energy for the CE phase, while for the Zener polaron phase the energy decreases linearly.  Hence, there is an instability towards the distortion of the lattice.

\begin{figure}[tb]
\includegraphics[width=0.37\textwidth]{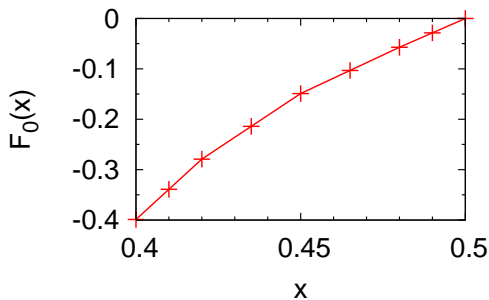}
\caption{\label{force}The nonzero value of $F_{0}(x)$ indicates an instability towards lattice distortions.  Only the CE phase does not present the instability. }
\end{figure}

In order to find the instability, we calculate the derivatives of the energy with respect to the distortions for the expansion
\begin{equation}
E=E_{0}+\frac{\partial E}{\partial \delta}\delta+O(\delta^{2}).
\end{equation}
The rate of change of the energy with respect to the distortion relates to the rise of ferroelectricity.  Introducing $\tilde{\delta}=\delta/a$, the energy we find has the form
\begin{equation}
E(\delta,x)=\left(F_{0}(x)\tilde{\delta}+E_{0}(x)\right)t,
\end{equation}
where $F_{0}(x)$ is a force, and $E_{0}(x)$ is the ground state energy without distortions.   The force, $F(\delta,x)$, is obtained by taking the derivative of the energy with respect to the distortion.  We find that $F_{0}(x)$ is the instability of the undistorted lattice towards lattice distortions.  The energies without the elastic energy contribution decrease linearly as a function of the distortions.  Thus, we conclude that the addition of the elastic energy will stabilize the system. 
As seen in Fig.~\ref{force}, the CE phase has no instability, while for the Zener polaron phase we see a large instability.  
 
Subsequently, we add the elastic energy term, $(1/16)D\delta^{2}$, where $D$ is the elastic constant, in order to find the minimum energy $E_{0}(\delta,x)$.  Defining $\tilde{D}=Da/16t$ the energy now has the form
\begin{equation}
E(\delta,x)=\left(\tilde{D}\tilde{\delta}^{2}+F_{0}(x)\tilde{\delta}+E_{0}(x)\right)t.
\end{equation}
Its minimum is given by the value of the distortion, $\delta_{0}$, for which $F(\delta_{0},x)=0$.  Taking the derivative of the energy with respect to the distortion we find 
\begin{equation}
\tilde{\delta}_{0}= -\frac{F_{0}(x)}{2\tilde{D}}.
\end{equation}
The elastic constant $\tilde{D}$ can be calculated for the Zener polaron phase with $\tilde{D}= -\tilde{F}_{0}(x)/2\tilde{\delta}_{0}$, with $\tilde{\delta}_{0}\approx0.042$ obtained from experimental results~\cite{exp}, where $a$ is the undistorted lattice spacing between sites.  Hence, we find a value of $\tilde{D}=16.62$, which we consider to be independent of the doping.  Using $t=0.622$ eV from Ref.~\cite{millis} and $a=3.947$ \AA~from Ref.~\cite{latcon} we recover $D=10.61$ eV/\AA$^{2}$, which is in agreement with other calculations~\cite{elastic}.

\begin{figure}[tb]
\includegraphics[width=0.43\textwidth]{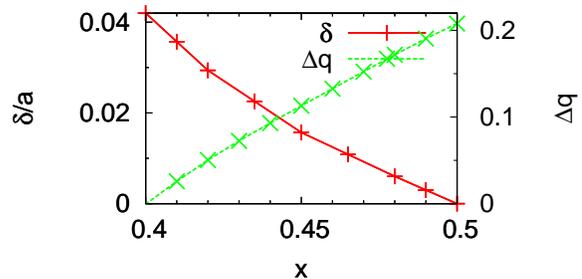}
\caption{\label{DeltaMin}Charge disproportionation (green), and lattice distortion (red) as a function of doping.  The Zener polaron phase has no dipole moment due to $\Delta q =0$, and the CE phase since $\delta=0$.}
\end{figure}

\begin{figure}[tb]
\includegraphics[width=0.395\textwidth]{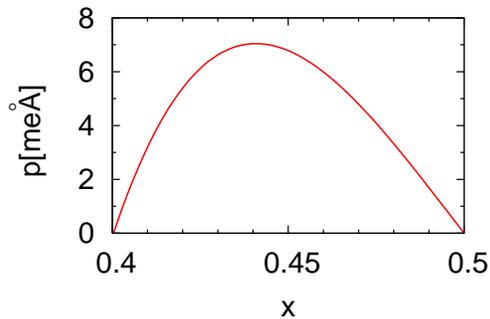}
\caption{\label{dm}Electric dipole moment $p$ as a function of doping $x$.   The maximum value of the dipole moment is $p_{max}\approx7.5$ me\AA ~at around $x\approx0.44$}
\end{figure}

Therefore, the dipole moment,
\begin{equation}
\vec{p}=\vec{\delta}\Delta q,
\end{equation}
is zero for the Zener polaron phase and the CE phase.  Nonetheless, the cause is different:  in the case of the CE phase, it is due to the lack of distortions, and for the Zener polaron phase due to a vanishing charge disproportionation, as shown in Fig.~\ref{DeltaMin}.  Strikingly, the in between phases have a nonzero electric dipole moment, shown in Fig.~\ref{dm}, that is directly proportional to the on-site Coulomb interaction:  $\Delta q$ increases as $U$ increases.  This counter intuitive result states that the on-site Coulomb interaction promotes a metallic state instead of localizing the electrons for dopings in between 0.4 and 0.5. 

\begin{figure}[tb]
\includegraphics[width=0.485\textwidth]{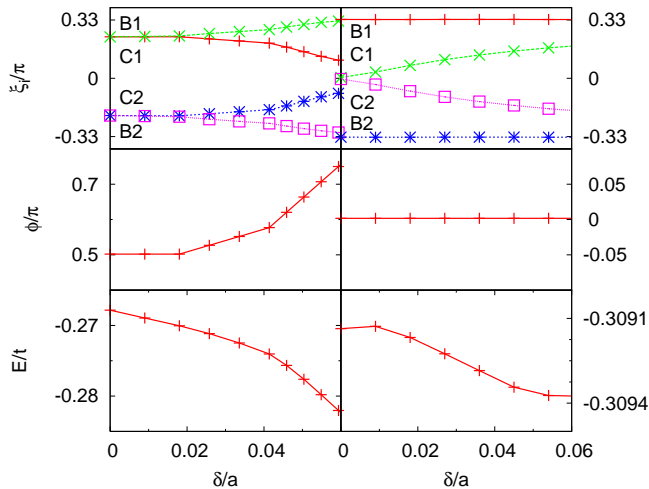}
\caption{\label{DistCant}Orbital order (top), spin canting (middle), and  energy without elastic energy contribution (bottom).  The left column is for the limiting case $x=0.4$ and the right for $x=0.5$.  For the range in which the energy changes linearly with the distortion the spin and orbital degrees of freedom do not change. }
\end{figure}

Finally, we extend our calculation to also include variations of the spin canting.  The result is shown in Fig.~\ref{DistCant} for the Zener polaron phase and CE phase.  For the CE phase we find that the spin canting is independent of the orbital and lattice distortion degrees of freedom.  The Zener polaron phase shows only a small range where the spin canting and the orbital order are independent of the lattice distortions.  Using again the lattice distortions $\tilde{\delta}\approx0.042$ for the Zener polaron phase, we see two effects:  firstly, the energy without the elastic energy decreases linearly with the distortions up to this value of $\delta$, and secondly, the orbital order and spin canting now have small corrections.  Interestingly, the effect on the orbital order is to make the two sites within a dimer slightly unequivalent.  For larger distortions that go beyond the stable point, the energy changes its dependence on the distortion, and we find a larger rate of change of the spin canting and orbital states of the underdoped states towards a CE phase configuration.  The elastic constant becomes $\tilde{D}=14.21$, which means that the energy of the phonons decreases with the spin canting.  In this case we recover $D=9.07$ eV/\AA$^{2}$.  For other dopings within the range here discussed we expect the same behavior.

We conclude that in underdoped manganites, such as La$_{1-x}$Ca$_{x}$MnO$_{3}$, there is an instability towards lattice distortions for the states in between 0.4 and 0.5 doping.  This gives rise to a dipole moment that results in a ferroelectric state.  Counterintuitively, stronger on-site Coulomb interactions result in a larger dipole moment, promoting ferroelectricity.  That charge order can exist in a ferromagnetic region has already been experimentally discussed~\cite{coinf}.  We have shown that the spin and orbit degrees of freedom are coupled to the distortion of the lattice.  It has been proposed that including the oxygens in the calculations results in the CE phase being made up of oxygen stripes~\cite{little}.  Interestingly, in our findings the magnetic correlation between dimers in the same stripe is always FM and does not change for the dopings being discussed.  We raise the question of whether these two results are related.  We do not find any agreement with the results in which a different configuration of the CE phase was proposed~\cite{orla}.  In addition, our results do not agree with the claim that Zener polarons are formed within the zig-zag chains of the CE phase and that they are important for understanding the localization of electrons in half-doped manganites~\cite{exp}.  Future work should include nonzero temperature calculations.

The author is deeply grateful to Frank Kr\"uger for lengthy discussions, and to Jens Bardarson for support while writing the code for the numerics.

This work was supported by the Stichting voor Fundamenteel Onderzoek der Materie (FOM), which is supported by the Nederlandse Organisatie voor Wetenschappelijk Onderzoek (NWO).

\end{document}